\documentclass[aps,prb,twocolumn,showpacs,superscriptaddress]{revtex4}

\usepackage{graphicx}
\usepackage{amsmath}

\newcommand{\rcg}{\textit{R}$_2$CoGa$_8$}
\newcommand{\tcg}{Tm$_2$CoGa$_8$}
\newcommand{\ecg}{Er$_2$CoGa$_8$}
\newcommand{\etal}{\textit{et al.}}
\newcommand{\sg}{$P4/mmm$}

\begin{document}

\title{The magnetic structures of the anisotropic inter-metallic compounds\\ \ecg\ and \tcg.}

\author{R. D. Johnson}\email{r.d.johnson@durham.ac.uk}
\author{T. Frawley}
\affiliation{Department of Physics, Durham University, South Road, Durham, DH1 3LE, UK}
\author{P. Manuel}
\author{D. D. Khalyavin}
\affiliation{ISIS Facility, Rutherford Appleton Laboratory-STFC, Chilton, Didcot, Oxon, OX11 0QX, UK}
\author{C. Adriano}
\author{C. Giles}
\author{P. G. Pagliuso}
\affiliation{Instituto de F$\acute{\textit{\i}}$sica Gleb Wataghin, Universidade Estadual de Campinas, 13083-970, Campinas - SP, Brazil}
\author{P. D. Hatton}
\affiliation{Department of Physics, Durham University, South Road, Durham, DH1 3LE, UK}

\date{\today}

\begin{abstract}

Two members of the iso-structural \rcg\ inter-metallic series, \ecg\ and \tcg, have been studied by powder neutron diffraction. Antiferromagnetic ordering of the rare-earth sublattices was confirmed to occur at  3.0~K and 2.0~K, respectively. Furthermore, determination of the critical exponent showed \ecg\ to adopt a 3D universality class. In spite of a common magnetic easy axis and similar structural characteristics, the antiferromagnetic structures were found to be different for the erbium and thulium based compounds. The corresponding magnetic space groups were determined to be $P_{2a}mmm'$ and $P_Cmmm$. The difference in magnetic structures is discussed based on crystal electric field effects that are known to be prevalent in such materials.

\end{abstract}


\maketitle

\section{Introduction\label{intro}}

Rare-earth inter-metallic compounds are a class of material that exhibit a diverse range of fascinating physical properties. For example heavy-fermion behaviour, Kondo ground states, quantum criticality, and pressure induced superconductivity have all been found.\cite{chen02,nicklas03,sugitani06,kimura07,bauer07} These phenomena are likely to be due to the competition of microscopic electronic interactions such as the magnetism and crystal electric field (CEF).\cite{pagliuso06,settai07,joshi07} A common theme in many intermetallics is the dependence of the electronic properties on the rare-earth ion. For example, CeCu$_2$Si$_2$ was found to undergo a transition into a superconducting state below $\sim$1~K that did not comply with the conventional Bardeen-Cooper-Schrieffer theory of superconductivity. However, LaCu$_2$Si$_2$ remains in a normal state down to 50~mK.\cite{steglich79} Further, a number of different magnetic structures have been measured across the series, solely dependent upon the choice of rare-earth ion.\cite{szytula94} The role of the rare-earth ion in the microscopic behavior of these systems remains an important question, specifically the interaction of the 4\textit{f} magnetism with the CEF. In this paper we concentrate on the newly synthezized \rcg\ series, which has been shown to exhibit a strong coupling between the 4\textit{f} magnetism and the CEF.\cite{joshi08} Consequently, a variety of electronic properties have been shown to be dependent upon the rare-earth ion.\cite{joshi08,joshi09}

The \rcg\ series is limited to \textit{R} = Gd~-~Lu and Y due to an instability in the crystallization of the lighter rare-earth compounds.\cite{joshi08,joshi09,adriano08} The series is iso-structural, adopting at room temperature the tetragonal space group \sg. This type of structure hosts a number of recently discovered heavy-fermion superconductors such as Ce$_2$CoIn$_8$,\cite{chen02} Ce$_2$RhIn$_8$,\cite{nicklas03} and Ce$_2$PdIn$_8$.\cite{kaczorowski09} The crystal structure can be thought of as a stacking of $R$Ga$_3$ units between CoGa$_2$ layers in the direction of the four-fold tetragonal axis. A thorough survey of magnetization and transport properties was published by Joshi \etal,\cite{joshi08,joshi09} in which an in-depth comparison between the \rcg\ series and other rare-earth inter-metallic compounds was presented. Of particular interest is the dependence of the magnetic anisotropy upon the rare-earth ion radii. This splits the series into four groups. These are the diamagnetic compounds with \textit{R}=Y and Lu, isotropic Gd$_2$CoGa$_8$ (as for Gd$^{3+}$ \textit{L}=0), \textit{R}=Tb, Dy and Ho compounds with a magnetization easy axis parallel to the \textit{c}-axis and finally \textit{R}=Er and Tm compounds, which have a magnetization easy axis perpendicular to the \textit{c}-axis. In this study we focus on the latter erbium and thulium based members. The two compounds antiferromagnetically order at $T_\mathrm{N}$~=~3.0 and 2.0~K, respectively,\cite{joshi08} and \ecg\ is on the border between having a magnetization easy axis parallel or perpendicular to the \textit{c}-axis.

In order to properly explain the bulk magnetic properties one must account for CEF effects. This is particularly apparent through CEF calculations that can explain the rare-earth dependent anisotropy of magnetization,\cite{joshi08} correctly predicting the direction of the magnetic easy axis and it's dependence upon rare-earth substitution. Furthermore, across the series, the antiferromagnetic transition temperatures are found to be higher than those predicted by de Gennes scaling. This can also be explained through CEF calculations.\cite{joshi08}

To better understand the varied macroscopic properties of inter-metallic materials we require a probe of the microscopic electronic ordering phenomena. The only previously published magnetic structure of any member of the \rcg\ series was that of Ho$_2$CoGa$_8$, determined by resonant magnetic x-ray scattering.\cite{adriano09} We have performed the first neutron diffraction study of this inter-metallic series. We show that \ecg\ and \tcg\ develop collinear antiferromagnetic structures below their respective N$\mathrm{\acute{e}}$el temperatures. In both materials, magnetic moments align parallel to the \textit{b}-axis, however they differ in propagation vector, \textbf{k}~=~(0,~1/2,~0) for \ecg\ and \textbf{k}~=~(1/2,~0,~1/2) for \tcg, due to CEF effects.

\section{Experiment\label{exp}}

Single crystal samples of \ecg\ and \tcg\ were grown by the gallium flux technique.\cite{fisk89} Lab-based x-ray powder diffraction confirmed the \ecg\ and \tcg\ room-temperature crystal structures to be tetragonal (\sg), with lattice parameters \textit{a}~=~4.210(5)~\AA\ and \textit{c}~=~10.96(1)~\AA, and \textit{a}~=~4.202(5)~\AA\ and \textit{c}~=~10.95(1)~\AA, respectively. Approximately 1~g of each sample was prepared for neutron powder diffraction by grinding selected single crystals using an agate pestle and mortar, resulting in fine powders of consistent grain size.

Neutron powder diffraction data were collected on both samples using the WISH time of flight instrument on the second target station at the ISIS facility.\cite{isis} A $^3$He sorption insert was employed within a standard Oxford Instruments cryostat to achieve sample temperatures of less than 300~mK. Each sample was loaded into a 6~mm diameter vanadium can with a thick copper head, covered with a Cd mask, placed in contact with the $^3$He pot. A copper wire (cold finger) was run through the length of the can to ensure better thermal conductivity through the sample. Data were collected with high counting statistics above $T_\mathrm{N}$ and at 300~mK, the base temperature of the $^3$He insert. Shorter data collections were performed upon warming through the transition to determine the temperature dependence of the magnetic scattered intensity. Determinations of the nuclear and magnetic structures were performed using the \textsc{fullprof} suite of programs.\cite{RodriguezCarvaja93}

\section{Results and Discussion\label{randd}}

The crystal structures of both compounds were refined above, and below, the respective magnetic transitions in the tetragonal \sg\ spage group, as has been reported for Ho$_2$CoGa$_8$.\cite{joshi08} In order to properly reproduce the experimental data, it was necessary to account for additional reflections due to extraneous scatter from the sample can. We therefore included in the refinements a copper nuclear phase (Cu cold finger). This phase was fitted by Le Bail intensity fitting as the copper wire was found to be extremely textured. In both samples we found no detectable changes of structural parameters above and below the magnetic transition, indicating that magneto-elastic coupling was negligibly small. We therefore only present the refinements of data measured below $T_\mathrm{N}$ at 300 mK. The lattice parameters and fractional coordinates of the inequivalent atom sites are presented in tables \ref{lattab} and \ref{crysttab}.

\begin{figure}
\includegraphics[width=8.5cm]{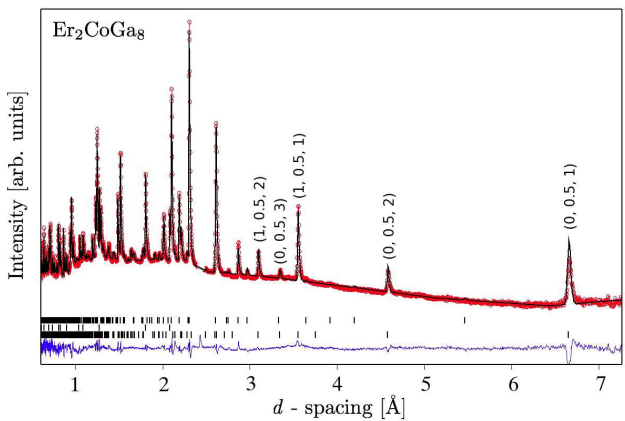}
\caption{\label{erdatafig}(Color online) Rietveld refinement pattern for \ecg\ at 300~mK. Only data from bank 3 (average $2\theta = 90^\circ$) of the WISH instrument was required to achieve high resolution over a sufficiently large time of flight interval to capture all the magnetic reflections. The top, middle and bottom tick marks refer to \ecg-nuclear, Cu-nuclear (cold finger) and \ecg-magnetic phases, respectively. The difference pattern is shown at the bottom of the figure (blue line). A number of prominent magnetic reflections are indexed.}
\end{figure}

\begin{figure}
\includegraphics[width=8.5cm]{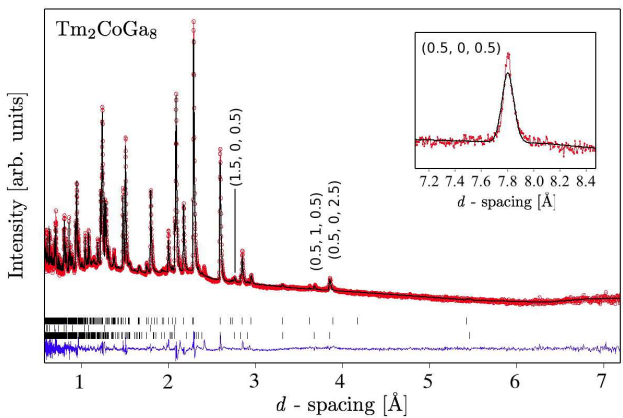}
\caption{\label{tmdatafig}(Color online) Rietveld refinement pattern for \tcg\ at 300~mK. Data from bank 3 (average $2\theta = 90^\circ$) of the WISH instrument is shown in the main pane, however it was necessary to incorporate data from bank 2 (average $2\theta = 58.33^\circ$) into the refinement in order to capture all high \textit{d}-spacing magnetic reflections (shown in inset). The top, middle and bottom tick marks refer to \tcg-nuclear, Cu-nuclear (cold finger), and \tcg-magnetic phases, respectively. The difference pattern is shown at the bottom of the figure (blue line). A number of prominent magnetic reflections are indexed.}
\end{figure}

\begin{figure}[t]
\includegraphics[width=8.5cm]{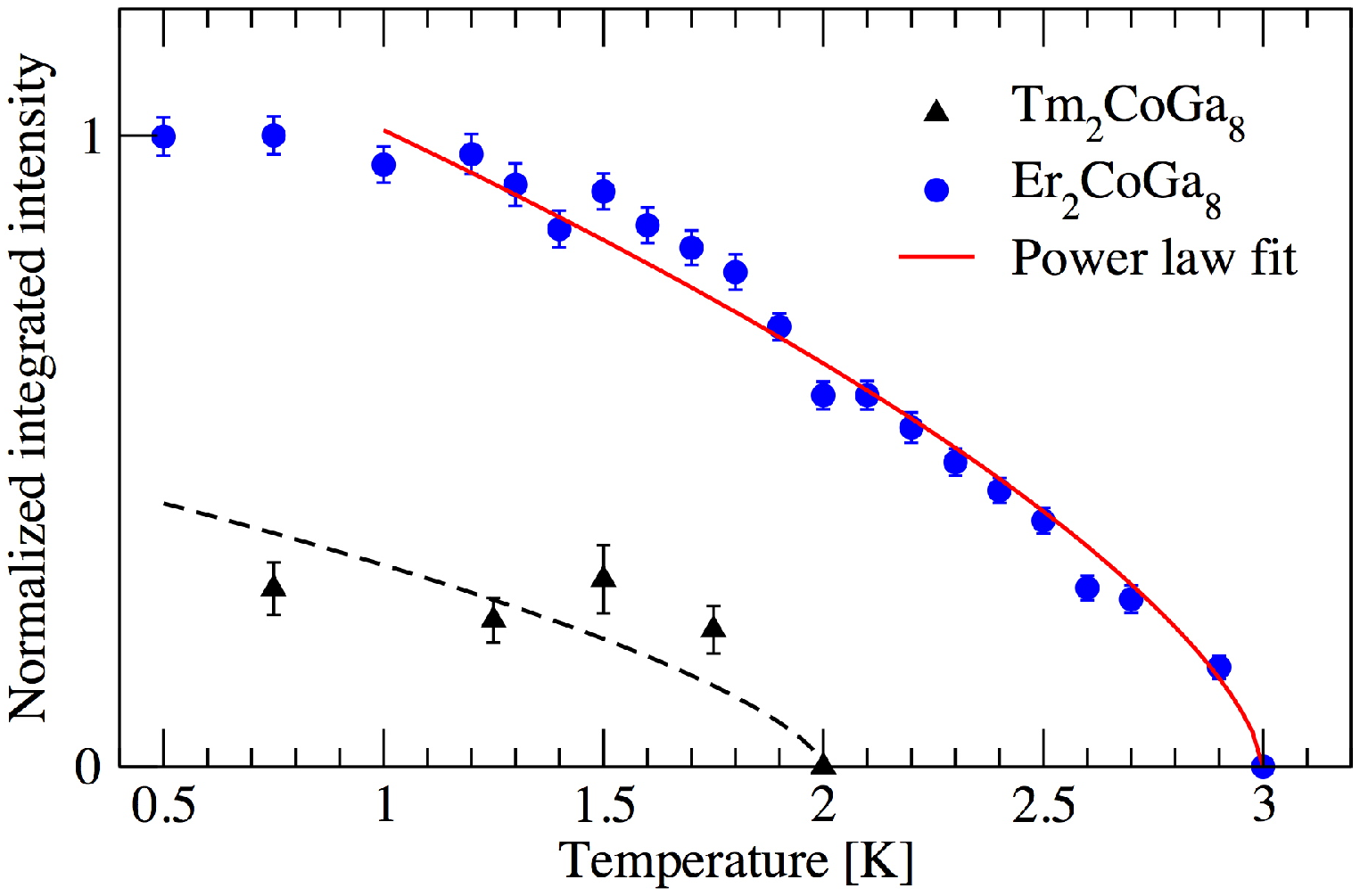}
\caption{\label{tempfig}(Color online) The temperature dependence of the integrated intensity of a selected magnetic reflection of both \ecg\ (\textit{d}~$\simeq$~3.56~\AA ) and \tcg\ (\textit{d}~$\simeq$~3.86~\AA ), shown as blue circles and black triangles, respectively. A power law has been fitted to the \ecg\ data (solid red line), giving a critical exponent of $\beta=0.33\pm0.02$ and $T_\mathrm{N}=3.00\pm0.01$ K. The \tcg\ data was not of sufficient quality to obtain a reliable fit, however the transition at $T_\mathrm{N}=2$ K is clear. A scaled power law with the same exponent as that fitted to the \ecg\ data is overlaid (broken black line).}
\end{figure}

Figures \ref{erdatafig} and \ref{tmdatafig} show the 300~mK neutron diffraction pattern and Rietveld refinements of both samples. By comparison between the powder patterns measured above $T_\mathrm{N}$ (not shown here) and at 300~mK, a large number of additional magnetic reflections became evident. We investigated the behaviour of the magnetic phase upon warming through the transition. The integrated intensites of selected magnetic diffraction peaks are plotted as a function of temperature in figure \ref{tempfig}, clearly showing transition temperatures of 3.0 and 2.0~K for \ecg\ and \tcg, respectively. This is in agreement with bulk magnetometry results.\cite{joshi08} Further, by fitting a power law to the \ecg\ data (the \tcg\ data is insufficient for fitting) we find a critical exponent of $\beta=0.33\pm0.02$, suggesting that these compounds adopt either the 3D-Ising or 3D-XY universality class.\cite{collins89} In either case we predict a three dimensional magnetic system, with 1D or 2D order parameters, respectively.

\begin{table}
\caption{\label{lattab}Lattice parameters refined from neutron powder data of \ecg\ and \tcg, measured at 300 mK.}
\begin{ruledtabular}
\begin{tabular}{c c c c}
Sample & Space group & \textit{a} (\AA) & \textit{c} (\AA) \\
\hline
\ecg & \sg & 4.20195(5) & 10.9438(2) \\
\tcg & \sg & 4.18980(7) & 10.9109(3) \\
\end{tabular}
\end{ruledtabular}
\end{table}

\begin{table}
\caption{\label{crysttab}Structural and magnetic parameters of \ecg\ and \tcg, refined from data collected at 300 mK. The direction of the magnetic moments of the rare-earth ions are chosen to lie parallel to the \textit{b}-axis (as opposed to the \textit{a}-axis).}
\begin{ruledtabular}
\begin{tabular}{c c c c c}
Atom & \textit{x} & \textit{y} & \textit{z} & Moment ($\mu_\mathrm{B} \Vert b$)\\
\hline
\multicolumn{5}{c}{\ecg, \textbf{k} = (0, 1/2, 0)}\\
Er (1) & 0 & 0 & 0.3068(3) & 4.71(3) \\
Er (2) & 0 & 0 & -0.3068(3) & -4.71(3) \\
Co & 0 & 0 & 0 & - \\
Ga (1) & 0 & 0.5 & 0.5 & - \\
Ga (2) & 0.5 & 0.5 & 0.2952(5) & - \\
Ga (3) & 0 & 0.5 & 0.1177(2) & - \\
\\
\multicolumn{5}{c}{\tcg, \textbf{k} = (1/2, 0, 1/2)}\\
Tm (1) & 0 & 0 & 0.2964(7) & 2.35(4) \\
Tm (2) & 0 & 0 & -0.2964(7) & -2.35(4) \\
Co & 0 & 0 & 0 & - \\
Ga (1) & 0 & 0.5 & 0.5 & - \\
Ga (2) & 0.5 & 0.5 & 0.3100(6) & - \\
Ga (3) & 0 & 0.5 & 0.1176(2) & - \\
\end{tabular}
\end{ruledtabular}
\end{table}

In the refinement of the magnetic structures, it was assumed that only the rare-earth ions are magnetic. The cobalt ions are expected to be non-magnetic due to an effective filling of the transition metal 3d states by an excess of gallium 4p electrons in the conduction band.\cite{wallace73} This is confirmed by the fact that Y$_2$CoGa$_8$ and Lu$_2$CoGa$_8$ are diamagnetic metals. Furthermore, the low transition temperatures are a signature of probable rare-earth ordering.

Despite having very similar crystal structures and common magnetic easy axes,\cite{joshi08} the 300~mK diffraction patterns (Fig. \ref{erdatafig} and \ref{tmdatafig}) clearly show that the two samples have different magnetic propagation vectors. These were determined to be associated with the X(\textbf{k}=(0,~1/2,~0)) and R(\textbf{k}=(1/2,~0,~1/2)) points of symmetry (Miller and Love notations\cite{miller67}) for the erbium and thulium compositions, respectively. In tetragonal symmetry, the choice of the \textit{a} and \textit{b} axes is arbitrary when defining the propagation vectors. The important consideration is the direction of the magnetic moments in the \textit{ab}-plane with respect to the propagation vector. Here, the \textit{a} and \textit{b} axes have been defined such that the magnetic moments lie parallel to the \textit{b}-axis, as will become apparent later. In both cases there are eight one-dimensional irreducible representations (irreps) associated with the corresponding wave vector groups. Six of them enter into the global reducible magnetic representation on the $2g$ Wyckoff position of the \sg\ space group occupied by Er or Tm. The symmetrized combinations of the axial vectors transforming as basis functions of these irreps correspond to the moment directions along the \textit{a}-, \textit{b}- and \textit{c}-axes with parallel and antiparallel alignment on the Er/Tm(1) and Er/Tm(2) sites.

Magnetic structure refinements corresponding to the six possible single-irrep models were performed for each sample. The magnetic and structural reliability factors (\textit{R}$_\mathrm{Mag}$ and \textit{R}$_\mathrm{Bragg}$) and $\chi^2$ for the refinements are given in Table \ref{rfactorstab}. It is clear that for each sample only a single model successfully fits the data, as highlighted in bold in Table \ref{rfactorstab} and shown in figures \ref{erdatafig} and \ref{tmdatafig}. We note that for all \tcg\ models, the $\chi^2$ values are similar. This is due to the magnetic reflections being weak compared to both the structural reflections and the background. However the \textit{R}$_\mathrm{Mag}$ values, which are also large due to the weak magnetic reflections, clearly show the correct model. For both \ecg\ and \tcg, the fitted magnetic structures correspond to moment directions aligned parallel to the \textit{b}-axis (with respect to our defined propagation vectors) and an AFM stacking along the \textit{c}-axis of ($+-+-$) and ($++--$), respectively. In a magnetic resonant x-ray diffraction study on Ho$_2$CoGa$_8$\cite{adriano09} it was not possible to determine whether the holmium moments stacked ($++--$) or ($+--+$) along the \textit{c}-axis. \tcg\ also adopts a similar \textit{c}-axis stacking, however our results are not ambiguous, and clearly show a single magnetic structure solution.

\begin{table}
\caption{\label{rfactorstab}Reliability factors of refinements of the six possible single-irrep magnetic structure models for both \ecg\ and \tcg. }
\begin{ruledtabular}
\begin{tabular}{c c c c c}
Moment & \textit{c}-axis &\textit{R}$_\mathrm{Mag}$ & \textit{R}$_\mathrm{Bragg}$ & $\chi^2$\\
axis & stacking & & &\\
\hline
\multicolumn{5}{c}{\ecg, \textbf{k} = (0, 1/2, 0)}\\
\textit{a} & ($++++$) & 67.5 & 11.7 & 43.4 \\
\textit{a} & ($+-+-$) & 68.7 & 9.05 & 29.3 \\
\textit{b} & ($++++$) & 75.7 & 11.0 & 42.4 \\
\textbf{\textit{b}} & \textbf{($\mathbf{+-+-}$)} & \textbf{14.7} & \textbf{7.29} & \textbf{8.91} \\
\textit{c} & ($++++$) & 83.6 & 11.7 & 44.3 \\
\textit{c} & ($+-+-$) & 45.5 & 9.7 & 20.0\\
\\
\multicolumn{5}{c}{\tcg, \textbf{k} = (1/2, 0, 1/2)}\\
\textit{a} & ($+--+$) & 102.0 & 8.97 & 14.3 \\
\textit{a} & ($++--$) & 54.4 & 7.5 & 12.1 \\
\textit{b} & ($+--+$) & 114.3 & 9.1 & 14.4 \\
\textbf{\textit{b}} & \textbf{($\mathbf{++--}$)} & \textbf{21.4} & \textbf{6.9} & \textbf{10.3} \\
\textit{c} & ($+--+$) & 93.9 & 9.0 & 14.4 \\
\textit{c} & ($++--$) & 50.8 & 7.7 & 12.5 \\
\end{tabular}
\end{ruledtabular}
\end{table}

It should be pointed out that the wavevector stars related to the X and R reciprocal points both consists of two arms. The corresponding non-collinear two-k magnetic structures are undistinguishable from the discussed above collinear single-k models in the powder diffraction experiments. The discrimination can be done only based on mono-domain single crystal measurements or based on observation of nuclear satellite reflections associated with the M(\textbf{k}=(1/2,1/2,0)) reciprocal point and having a specific critical behavior. These reflections are expected in the case of the two-k structures due to the presence of appropriate coupling invariants in a polynomial decomposition of the Landau free energy. The observation of these extremely weak reflections is a difficult experimental task and is possible only in single crystal measurements. However, taking into account the simple exchange topology of the rare-earth sub-lattice in \rcg, without geometrical frustration, the non-collinear two-k structures are considered here to be unlikely.

The magnitudes of the magnetic moments at 300~mK were refined to be 4.71(3)~$\mu_\mathrm{B}$/Er and 2.35(4)~$\mu_\mathrm{B}$/Tm, as given in Table \ref{crysttab}. These values are in good agreement with those obtained from bulk magnetization measurements; extrapolating the isothermal magnetization data measured by Joshi \etal\cite{joshi08} gives zero field magnetic moments of $4.6~\pm0.1~\mu_\mathrm{B}/\mathrm{Er}$ and $2.90~\pm0.03~\mu_\mathrm{B}/\mathrm{Tm}$. The rare-earth ion moments are found to be much smaller than their theoretical free-ion values of 9~$\mu_\mathrm{B}$/Er and 7~$\mu_\mathrm{B}$/Tm. This is expected in such systems where the magnetic behaviour is dominated by crystal electric field (CEF) effects.\cite{szytula94}

The ground state multiplet degeneracy of the \textit{R}$^{3+}$ ions is lifted by the CEF. The wave functions of the split energy levels were calculated from the CEF parameters found by Joshi \textit{et al}.,\cite{joshi08} in terms of the basis states $|$J,J$_\mathrm{z}\rangle$, using the McPhase software package.\cite{mcphase} The zero-field ground states of both Er$^{3+}$ (doublet) and Tm$^{3+}$ (singlet) are non-magnetic. The state energies and associated $\langle\mathrm{J}_\mathrm{z}\rangle$ were calculated as a function of internal magnetic field. For Er$^{3+}$, $\langle\mathrm{J}_\mathrm{z}\rangle=4.0~\mu_\mathrm{B}$ in the field range 18 to 47~T, and for Tm$^{3+}$, $\langle\mathrm{J}_\mathrm{z}\rangle=2.7~\mu_\mathrm{B}$ in the field range 1.5 to 76~T. These values, clearly illustrating the reduction in moment due to the CEF, are close to those determined from neutron powder diffraction. The deviation from the empirically determined values is likely to be due to uncertainties in the CEF parameters, upon which the calculation is based.

\begin{figure}
\includegraphics[width=7cm]{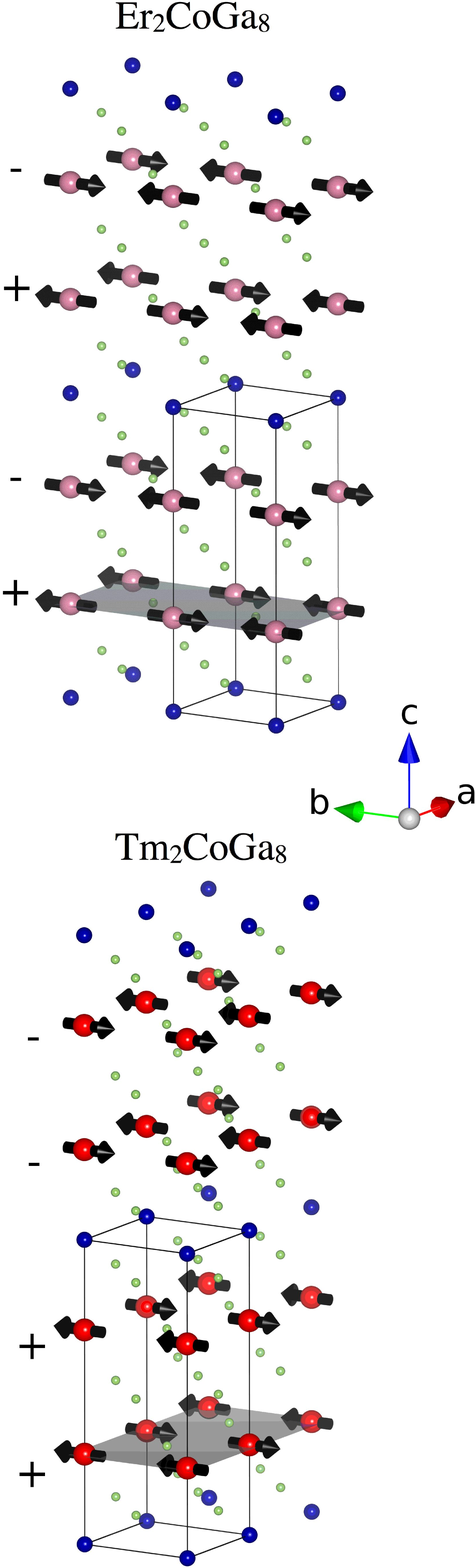}
\caption{\label{strucfig}(Color online) The crystal structure of \ecg\ (top) and \tcg\ (bottom) with the respective magnetic structures superimposed, represented by black arrows. Erbium, thulium, cobalt and gallium ions are shown by pink, red, blue and green spheres of decreasing size, respectively. The \sg\ unit cell is drawn in black, and \textit{ab}-planes of rare-earth ions are shaded grey.}
\end{figure}

In \ecg\ magnetism propagates antiferromagnetically (AFM) along the \textit{b}-axis in the direction of the moments, which ferromagnetically (FM) couple along the \textit{a}-axis (figure \ref{strucfig} top). The \textit{ab}-planes of rare-earth ions, containing the easy axis of magnetization, are stacked AFM ($+-+-$) along the \textit{c}-axis. The magnetic space group is $P_{2a}mmm'$ (No. 355)\cite{litvin08} with lattice vectors being (0,2,0), (0,0,1) and (1,0,0) with respect to the basis of the parent \sg1$'$ gray group. By contrast, in \tcg\ the moments in the \textit{ab}-plane align in an opposite fashion to \ecg\ \textit{i.e}. FM along the \textit{b}-axis and AFM along the \textit{a}-axis (figure \ref{strucfig} bottom). Furthermore, the planes are stacked alternately FM / AFM along the \textit{c}-axis ($++--$). This magnetic structure can be described by the $P_Cmmm$ (No. 353)\cite{litvin08} space group with (2,0,0), (0,0,-2) and (0,1,0) basis vectors with respect to \sg1$'$.

The above magnetic space groups are orthorhombic, and as such the tetragonal symmetry of the crystal structure in the paramagnetic phase must be broken below $T_\mathrm{N}$. There was no evidence in our powder diffraction data of a lowering of crystal symmetry, however the crystallographic distortions due to the magneto-striction are expected to be extremely small (of the order 10$^{-4}$ to 10$^{-6}$); beyond the instrument resolution. The crystal structure was therefore refined in tetragonal symmetry below $T_\mathrm{N}$, as given above, despite the symmetry lowering. In this scenario the magnetic structure with \textbf{k}=(0,~1/2,~0) and moments parallel to the \textit{b}-axis (in the case of \ecg) is exactly equivalent to a structure with \textbf{k}=(1/2,~0,~0) and moments parallel to the \textit{a}-axis. In fact, these cases correspond to two different domains associated with the different arms of the same wavevector star. Indeed, refinements of both magnetic structures gave equivalent \textit{R}$_\mathrm{Mag}$, \textit{R}$_\mathrm{Bragg}$ and $\chi^2$.

The spacing of the rare-earth ions is approximately 10 times greater than the typical radius of the localized, magnetically ordered rare-earth 4f states ($\sim$0.5~\AA). There is insufficient overlap of wavefunctions for direct exchange to occur. The dominant exchange interaction in these compounds is therefore the Ruderman-Kittel-Kasuya-Yosida (RKKY) interaction that supports both FM and AFM exchange, dependent upon the inter-atomic distances. In competition with the RKKY exchange integral are the CEF terms of the Hamiltonian.\cite{szytula94} As a consequence, rare-earth inter-metallic systems may exhibit a wide variety of magnetic structures ranging from collinear FM\cite{routsi92} to spin glasses.\cite{goremychkin08} As both \ecg\ and \tcg\ have similar inter-atomic distances between the rare-earth ions, we suggest that the difference in magnetic structures is due primarily to the CEF terms. Joshi \etal\cite{joshi08} showed that the extreme magnetic anisotropy exhibited by this inter-metallic series can be accounted for by CEF effects. The magnetic structures refined for \ecg\ and \tcg\ clearly give easy axes lying in the \textit{ab}-plane. The CEF results in a higher susceptibility in the \textit{ab}-plane, and the magnetic structure allows for a canting of moments in the same direction. Furthermore, in their calculations of the CEF, they find the in-plane and \textit{c}-axis nearest-neighbor exchange constants to be $\mathcal{J}_{ex}^{ab}/k_B=-0.35$~K and $\mathcal{J}_{ex}^{c}/k_B=-0.32$~K for \ecg, and  $\mathcal{J}_{ex}^{ab}/k_B=-0.53$~K and $\mathcal{J}_{ex}^{c}/k_B=-0.075$~K for \tcg.

In \ecg\ $\mathcal{J}_{ex}^{c}$ is  approximately equal to $\mathcal{J}_{ex}^{ab}$. Our results show that this scenario favors a ($+-+-$) AFM stacking of planes, with an equivalent, anisotropic AFM coupling in the \textit{ab}-plane, along the \textit{b}-axis. By comparison, in \tcg\ $\mathcal{J}_{ex}^{c}$ is small compared to the large $\mathcal{J}_{ex}^{ab}$ value. The magnetic structure is therefore dominated by a strong, in-plane AFM coupling with a favored ($++--$) \textit{c}-axis stacking. Our refined magnetic structures are consistent with, and reinforce, the calculated CEF dependent exchange constants. Importantly, this shows that by choice of the rare-earth ion one can modify the CEF within the \rcg\ series in order to predetermine the magnetic structure.

The \rcg\ crystal structure can be thought of as a stacking of \textit{R}Ga$_3$ units, separated by CoGa$_2$ layers. By comparing \textit{R}Ga$_3$ to \rcg, the similar \textit{a} lattice parameter and the requirement of heavier rare-earth mass for stable crystallization suggests that the \textit{R}Ga$_3$ units are key building blocks in the formation of the ternary compound.\cite{joshi08} An analogous argument is also made for \textit{R}$_2$CoIn$_8$\cite{joshi08,joshi07} and Ce$_2$RhIn$_8$\cite{bao01} compounds. Indeed, in Ce$_2$RhIn$_8$ this is supported by the common rare-earth magnetic structures of Ce$_2$RhIn$_8$ and CeIn$_3$.\cite{bao01} This suggests that the RhIn$_2$ layers have little influence on the magnetic structure, giving 2D characteristics. We note that this is not the case in the \rcg\ series. The simple collinear magnetic structures refined in this paper do not reflect the more complicated magnetic structures of the \textit{R}Ga$_3$ (\textit{R}~=~Er and Tm) compounds.\cite{morin87} For example, the TmGa$_3$ magnetic structure is multiaxial, involving two or three propagation vectors. Indeed, the evaluation of the critical exponent predicts a magnetic structure that is 3-dimensional. Further work is required to understand the role of the CoGa$_2$ layers and their counterparts in the other inter-metallics, particularly when considering the dimensionality of the magnetic structure.

\section{Conclusions\label{conc}}

We have solved the magnetic structure of two, newly synthezized inter-metallic compounds, \ecg\ and \tcg.  In both materials, magnetic moments were refined and found to lie parallel to the \textit{b}-axis, with magnitudes 4.71(3)~$\mu_\mathrm{B}$/Er and 2.35(4)~$\mu_\mathrm{B}$/Tm. Despite having common easy axes of magnetization (in the \textit{ab}-plane) the magnetic propagation vectors were found to be different; in \ecg\ \textbf{k}~=~(0,~1/2,~0) and in \tcg, \textbf{k}~=~(1/2,~0,~1/2). The different magnetic structures are due to a competition between crystal electric field effects and the RKKY exchange interaction. We show that the magnetic order parameter adopts either the 3D-Ising or 3D-XY universality class, with transition temperatures of 3.0~K and 2.0~K, for the erbium and thulium compounds, respectively. Further, by comparison of the \rcg\ and \textit{R}Ga$_3$ magnetic structures, we suggest that the CoGa$_2$ layers play an important role in the 3D magnetism in this series, as opposed to inducing a quasi-2D magnetic structure as in Ce$_2$RhIn$_8$.

\begin{acknowledgments}
RDJ would like to thank Stewart Bland for helpful discussions. RDJ, TF and PDH would like to thank STFC (UK) and EPSRC (UK) for funding. CA, CG and PGP acknowledge support from FAPESP (SP-Brazil), CNPq (Brazil) and CAPES (Brazil).
\end{acknowledgments}

\bibliography{r2coga8}

\end{document}